# mTOF performance during mCBM beam time at GSI

Q. Zhang[a,][-] I. Deppner,[b] N. Herrmann[b] and Y. Wang[a]

[a] *Key Laboratory of Particle and Radiation Imaging, Department of Engineering Physics, Tsinghua University*
  *Beijing 100084, China*
[b] *Physics Institute, Heidelberg University*
  *Heidelberg 69120, Germany*
  *E-mail*: zhangqn17@tsinghua.org.cn

ABSTRACT: The future Facility for Anti-proton and Ion Research (FAIR), currently in construction in Darmstadt, Germany, is one of the largest research projects worldwide. The Compressed Baryonic Matter (CBM) experiment is one of the main pillars at FAIR, studying the quantum chromodynamics (QCD) phase diagram at high baryon densities with unprecedented interaction rate in heavy ion collisions up to 10 MHz. This requires new data-driven readout chain, new data analysis methods and high-rate capable detector systems. The task of the CBM Time of Flight wall (CBM-TOF) is the charged particle identification. Multi-gap Resistive Plate Chambers (MRPCs) with different rate capabilities will be used at CBM-TOF corresponding regions. To reduce the commissioning time for CBM, a CBM full system test-setup called mini-CBM (mCBM) had been installed and tested with beams at GSI SIS18 facility in 2019. The high-rate MRPC prototypes developed at Tsinghua University, called MRPC2, were selected to be implemented in mTOF modules for mCBM. Additional thin float glass MRPCs from USTC called MRPC3, foreseen for the CBM lower rate region, were also tested in the mCBM experiment. Performance results of the two kinds of MRPCs analyzed by the so called tracking method will be shown.

---

[-] Corresponding author.

# Contents

## 1. Introduction

The future Facility for Anti-proton and Ion Research (FAIR) is currently under construction in Darmstadt, Germany. As a large and complex accelerator facility, it will be one of the largest research projects worldwide by providing particle beams of all chemical elements (or their ions), as well as anti-protons. The layout of FAIR is presented in Figure 1 left part [1]. As shown in Figure 1, the fixed target experiment Compressed Baryonic Matter (CBM) is one of the main pillars at FAIR, studying the quantum chromodynamics (QCD) phase diagram at high baryon densities with unprecedented interaction rate in heavy ion collisions up to 10 MHz. In order to measure the flight time of particles for hadron identification in CBM, a time-of-flight (TOF) wall called CBM-TOF is designed. CBM-TOF is composed of Multi-gap Resistive Plate Chambers (MRPCs) with a total active area of 120 $m^2$. It will be located downstream of the target as shown in the right part of Figure 1. The distance from CBM-TOF to the target is adjustable from 6 m to 10 m for different physics topics [2]. MRPCs with different rate capabilities are needed to be located at different areas of CBM-TOF according to the rate requirement. Since CBM is a high-rate experiment, the MRPCs that are closer to the beam are required to have a very high rate capability up to 30 kHz/$cm^2$, which is a challenge for MRPC detectors. This is because the rate capability of common float glass MRPC is only few hundreds Hz/$cm^2$ [3], which does not meet the requirement of CBM-TOF higher rate region where the rate is higher than 1 kHz/$cm^2$. Thus, new MRPCs with higher rate capability are designed.

  In addition, new readout concepts together with new data analysis methods are also needed and developed for the high-rate experiment. In order to reduce the commissioning time of CBM, a CBM full system test-setup called mini-CBM (mCBM) has been installed and operated with test beams at GSI SIS18 [4]. It is not only meant to test the performance of new counters but also new readout concepts, combining all subsystems, together with new data analysis methods under radiation conditions similar to the ones expected for CBM. The location of mCBM can be found in Figure 1, left part.

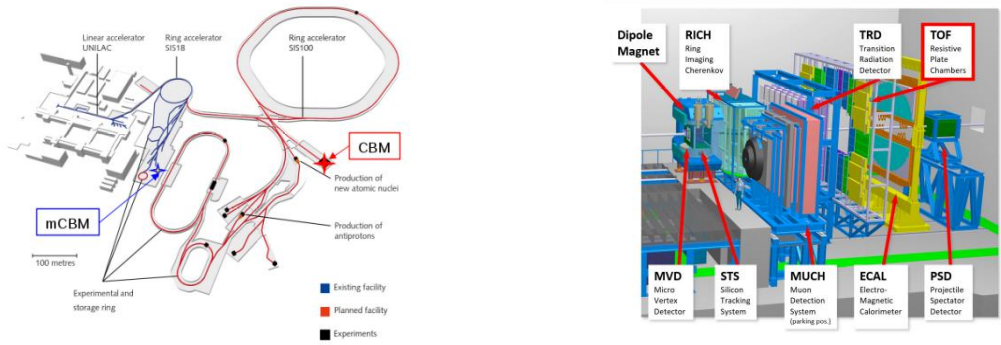

**Figure 1.** Left: the layout of FAIR where the red star represents the location of CBM and the blue star represents the location of mCBM. Right: the CBM experimental setup.

## 2. Experimental setup of mTOF

### 2.1 Structures of tested MRPC2 and MRPC3

According to the "Static" Model, the high rate capability of MRPC can be achieved either by reducing the resistivity of resistive plates or/and reducing the thickness of resistive plates [5]. In CBM-TOF project, the developments of two kinds of MRPC are close to be finished: a low resistive glass MRPC [6] called MRPC2 which is required to have a rate capability of 5 kHz/cm$^2$ and an ultra-thin float glass MRPC called MRPC3 which is required to have a rate capability of 1 kHz/cm$^2$ [7]. The structure of MRPC2 is shown by the schematic of the cross section in Figure 2. MRPC3 has a similar structure.

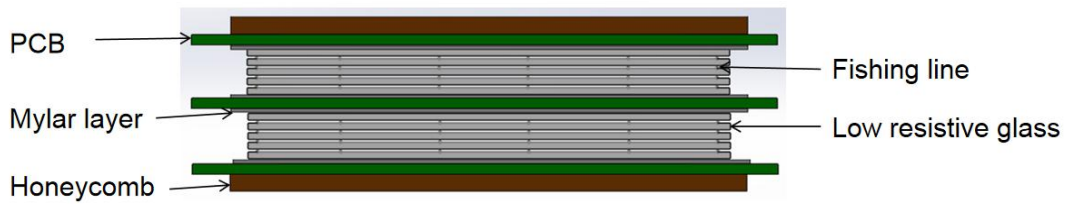

**Figure 2.** The cross section of MRPC2. MRPC3 has similar structure.

MRPC2 has 2×4 250 μm wide gas gaps while MRPC3 has 2×5 230 μm wide gas gaps [8]. The detailed parameters of the two kinds of MRPCs are summarized in Table 1. The main difference between the two MRPCs is the material of the resistive plate, which plays an important role for the rate capability. MRPC2 is equipped with low resistive glass with a bulk resistivity about 10$^{10}$ Ωcm. This resistivity is about two magnitudes smaller than that of common float glass [9]. MRPC3 uses ultra-thin float glass sheets as resistive plates to optimize its rate capability. 25 MRPC2 and 2 MRPC3 prototypes were tested at mCBM experiment.

**Table 1**. The parameters of MRPC2 and MRPC3

| Parameters of MRPC | MRPC2 | MRPC3 |
|---|---|---|
| Number of gas gaps | 8 (4 in each stack) | 10 (5 in each stack) |
| Gas gap width | 0.25 mm | 0.23 mm |
| Electrodes | Low resistive glass | Ultra-thin float glass |
| Thickness of glass electrodes | 0.7mm | 0.28mm |
| Strip length | 270 mm | 270 mm |
| Strip width | 7 mm | 7 mm |
| Strip pitch | 10 mm | 10 mm |
| Strip number | 32 | 32 |

## 2.2 The data-driven readout chain for MRPCs

Instead of the conventional triggered system, a new data-driven readout chain has been developed for all detector subsystems in CBM to meet the requirements from high-rate environment. The main feature is that self-triggered front-end electronics (FEE) delivers time-stamped data messages on activation of the respective detector channel [10]. The data sent by the FEE will be aggregated and transported to a high-performance online computer farm located in the Green IT Cube for data reconstruction and selection in real time [11]. The components of the available readout chain for the mCBM beam time in 2019 are shown in Figure 3. The pre-amplifier discriminator (PADI X) has 32 channels on each board and the threshold can be set via slow control. The PADI boards are directly connected to the MRPC readout electrodes inside the gas box. The advantages of this configuration are the suppression of the electromagnetic induced noise from outside and the impedance matched transmission line from electrodes to the pre-amplifier itself. The LVDS signals generated by PADI contains the time information in the leading edge and time over threshold (TOT) information in the signal width. Those signals are digitized by the GSI event-driven TDC (GET4) [12]. The GET4 boards are directly plugged to the feed through PCB in order to avoid cables outside the module. On the opposite side a backplane PCB with a radiation hard ASIC called GBTx collects and combines data from 40 GET4 chips (5 boards), sending them to a FPGA based Data Processing Board (DPB), which sits outside of the cave. The DPB concentrates the data from several GBTx boards and implements them into a data container with specific time interval. This process is called μ-time slice building. The μ-time slices from several DPBs are sent to the First Level event selector Interface Board (FLIB) where they are combined into time slices and stored on the hard drive for later offline analysis.

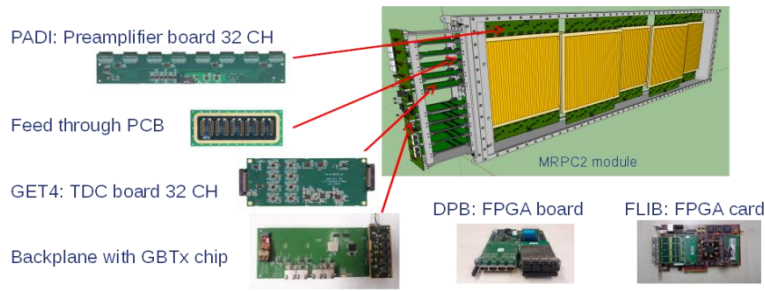

**Figure 3.** The readout chain for the 2019 mCBM beam time. It contains PADI, GET4, GBTx, DPB and FLIB.

## 2.3 The mTOF experimental setup at mCBM

The two kinds of MRPCs described in chapter 2.1 and other MRPC prototypes composed the TOF system at mCBM. During the 2019 beam time, the TOF system was placed at a distance of 234 cm from the target under an angel of about 25 degree with respect to the primary beam (c.f. Figure 4). The beam was composed of Ag ions impinging on two kinds of targets: a thin Au target with a thickness of 0.25 mm (about 1% interaction probability) or a thick target with a thickness of 2.5 mm (about 10% interaction probability). The system of 25 MRPC2 counters are forming the mini-TOF (mTOF) modules of mCBM. They are arranged into 5 mTOF modules housing 5 MRPC2s each. The modules are positioned in two different columns along the z direction with 3 layers at the 25 degree angle and 2 layers at the smaller angle (see Figure 4). The two additional MRPC3 counters are mounted together in a smaller gas box located behind the double layer stack. The other prototype counters have their own gas box and they are fixed at the back side of the setup. The double layer stack as well as the MRPC3 module is located closer to the beam pipe in order to expose them to a higher particle flux in comparison to the triple layer. With the multi-layer arrangement, the particle tracks passing through the mTOF setup can be reconstructed by combining the hit information on each layer. They will be used to evaluate the performance of the MRPCs. The working gas mixture is the standard mTOF gas mixture containing 90% $C_2H_2F_4$, 5% $i-C_4H_{10}$ and 5% $SF_6$, flowing through the modules in parallel with a total flux of 110 ml/min for the beam test.

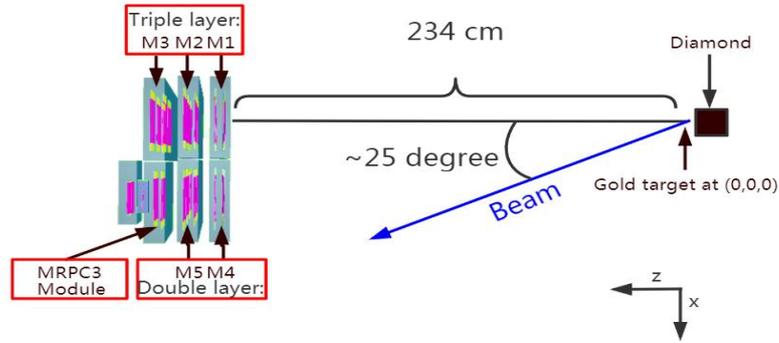

**Figure 4.** The mTOF experimental setup at mCBM. M1 represents the first MRPC2 module. The other MRPC2 modules are named accordingly.

## 3. Analysis and preliminary results of mCBM beam tests

The analysis is based on subset of data from the 2019 beam time specifically on data from run 159, where the beam of 1.58 A GeV Ag ions hit the thin Au target. The beam intensity was around $2 \times 10^5$ ion /s. The 25 MRPC2s and the 2 MRPC3s were set to the high voltage (HV) of $\pm$ 5300 V (corresponding to 106 kV/cm) and $\pm$ 6400 V (corresponding to 111.3 kV/cm), respectively. The difference of the electric field strength of MRPC2 and MRPC3 is motivated by the different gap size. This is confirmed by comic tests and previous spot beam test [13].

A tracking method is developed to evaluate the performance of the MRPC counters. The first step is to calibrate the raw experimental data. This includes the correction of position offset, time offset and walk. The next step is to form tracks from the diamond detector to the MRPCs. Each MRPC layer including additional MRPC prototypes is used as one tracking station. The efficiency is evaluated by comparing tracks where all stations have a hit $N_{Hit}$ with tracks where the counter under test is missing $N_{Miss}$ (c.f. Figure 5 left part). The right part shows the results, where the obtained global efficiency of an exemplary MRPC2 (center counter in second module) is above 91% as shown in upper right corner. The greenish band represents an area of low efficiency, that reduces the global efficiency. The band with low efficiency indicates that one GET4 chip (4 channels) was partially inactive due to loss of synchronization. The true value of global efficiency of the counter can be calculated by ignoring the low efficiency band amounting in about 95%. Out of 1600 channels from all 25 MRPC2s only one channel is broken. In the lower right corner of Figure 5 the efficiency distribution of an exemplary MRPC3 is shown. The global efficiency is 92.3%. Compared to MRPC2, the area available for efficiency calculation for MRPC3 is smaller. The reason is that the acceptance for tracking is limited by a small MRPC prototype with an active area of $18 \times 18$ cm$^2$ behind MRPC3.

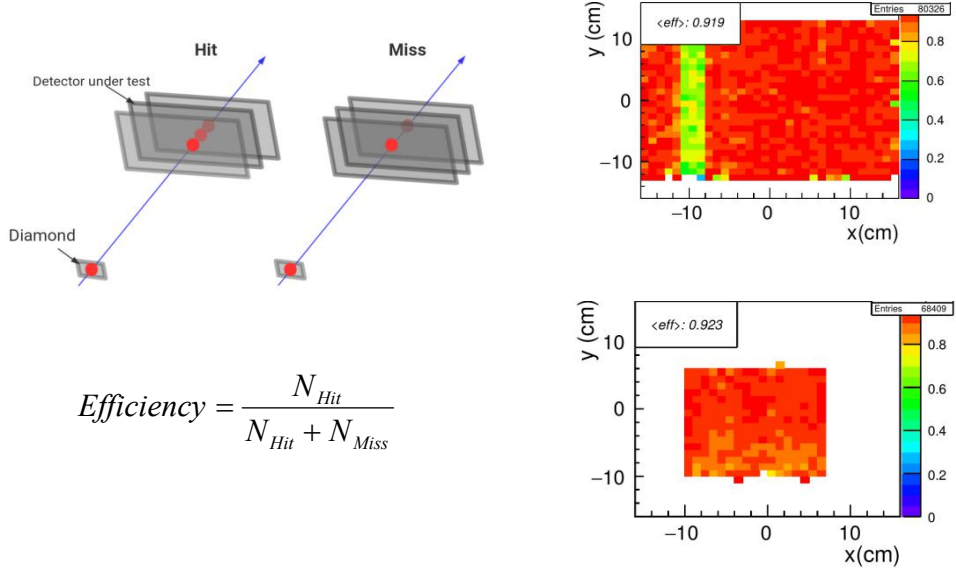

$$Efficiency = \frac{N_{Hit}}{N_{Hit} + N_{Miss}}$$

**Figure 5.** Left: the explanation of efficiency calculation based on tracking method. Right: an example of MRPC2's efficiency is on the upper side and an example of MRPC3's efficiency is on the lower side.

The time difference distribution based on tracking method can be deduced from the difference between the measured time and the projected time obtained from the fitted track where the fit does not include the measured hit. Figure 6 shows the time difference distributions for MRPC2 and MRPC3. The observed widths are about 87.4 ps and 62.9 ps, respectively, and are upper limits of the real MRPC timing resolution. The difference in the shown time distributions (see Figure 6) can be explained by the fact that in case of MRPC3, two more tracking stations contribute to the reference time.

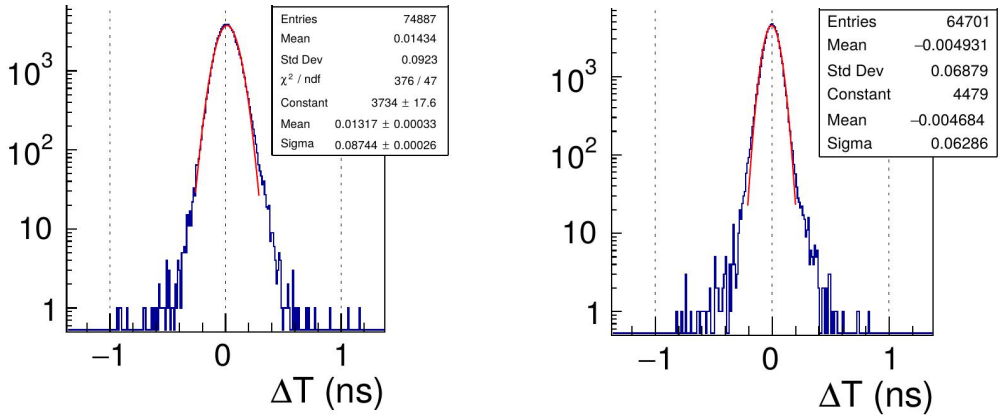

**Figure 6.** Left: the time difference distribution related with the exemplary MRPC2. Right: the time difference distribution related with the exemplary MRPC3 (right).

An interesting feature of the tracking method is the possibility to visualize different vertices as depicted in Figure 7 left part. Here, tracks neglecting the diamond counter are projected to the target plane shown in Figure 7 right part. The main contribution comes from the target and the diamond counter while the right spot is generated by the beam exit window. A Gaussian fit on the projected histogram delivers a pointing accuracy in the order of 4 cm to the target plane.

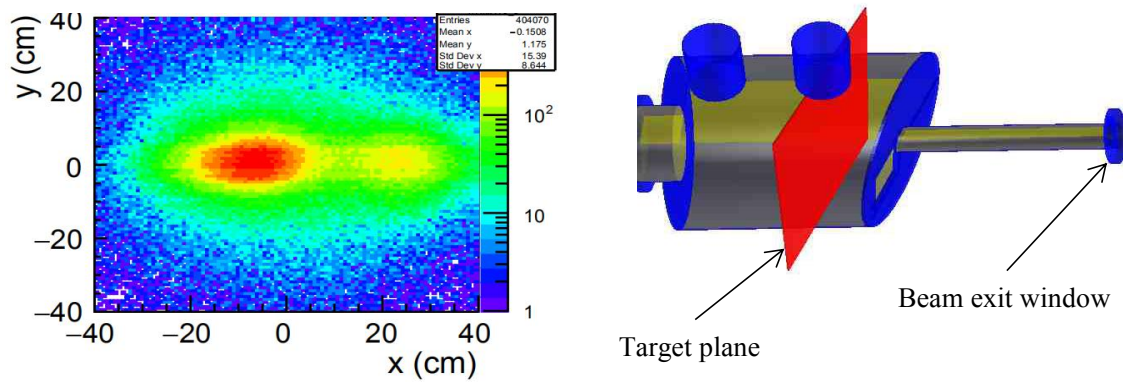

**Figure 7.** Left: the 2D target plane reconstructed by tracks among MRPCs. Right: target chamber with target plane shown in the red area.

## 4. Conclusion and outlooks

The MRPC2 and MRPC3 prototypes were tested with full illumination condition at mCBM in 2019. The beam test results prove the feasibility of data-driven readout chain together with tracking analysis method. Efficiency above 90% and time resolution better than 90 ps are achieved. However, more improvements related with DAQ firmware are needed and this is ongoing. The unfinished high-rate tests and aging tests with an improved firmware are scheduled for 2020 to evaluate the precise performance of CBM-TOF counters under CBM high rate running conditions. A more accurate efficiency determination of the counters can be expected when the readout problems are solved. In the mCBM, the first steps towards combined data taking were done and improvements are ongoing aiming at finally measuring the production probability of rarely produced Lambda baryons at SIS18 energies.

## Acknowledgments

This work is supported in part by BMBF 05P19VHFC1 and National Natural Science Foundation of China under Grant No.11420101004, 11461141011, 11275108, 11735009, 11927901. It is also supported by the Ministry of Science and Technology (China) under Grant No. 2015CB856905, 2016 YFA0400100, 2018YFE0205203.